\documentclass{article}
\usepackage{epsfig}
\usepackage{amssymb}
\usepackage{amsmath}
\usepackage{amsfonts}
\newcommand \be{\begin{eqnarray}}
\newcommand \ee{\end{eqnarray}}
\begin{document}
\begin{center}
{\bf The Triton; Low-momentum Interactions and 0ff-shell Effects}\\
\bigskip
\bigskip
H. S. K\"ohler \footnote{e-mail: kohler@physics.arizona.edu} \\
{\em Physics Department, University of Arizona, Tucson, Arizona
85721,USA}\\
\end{center}
\date{\today}

\begin{abstract}
A microscopic theory of nuclei  based on a 'free' scattering 
NN-potential is meaningful only if this potential fits on-shell
scattering data. 
This is a necessary
but not sufficient condition for the theory to be successful.
It has been demonstrated repeatedly in the past
that 2-body off-shell adjustments \it or \rm many-body forces are necessary. 
It has been shown however, using
Effective Field Theory (EFT) as well as formal scattering theory, that 
off-shell and many-body effects can not be separated.
This 'equivalence theorem' allows us to concentrate on the off-shell effects.
On-shell equivalent potentials can be constructed using
meson-theory (Paris, Bonn etc) but in this report 
separable potentials
are calculated  by inverse scattering from
NN-scattering and Deuteron data without any external parameters. Earlier
calculations showed these $S$-state potentials in agreement with  Bonn-B
results in Brueckner nuclear matter calculations. They are here also used to compute the 
Triton binding energy and the n-D  scattering
length $^2a$.
The results are found to lie on the Phillips line
defined in early calculations but like these miss
the experimental point on this line and overbind the Triton.
The point \it is \rm reached by modifying the off-shell properties accomplished
by adding a short-range repulsion without affecting fits to the
experimental low-energy  phase-shifts, i.e. the low energy \it free \rm two-body interaction. 
The off-shell induced correlations result in  a repulsive component in the Triton effective interactions.
In nuclear matter the same effect is referred to as the dispersion correction, which is a main contributor to
nuclear saturation.  In finite nucleus Brueckner-Hartree-Fock calculations these same correlations give an
important contribution to the selfconsistent field referred to as a reaarangement term, without which the
finite nucleus would collapse.
The main purpose of the present work is to illustrate that 
NN-correlations are as important in the Triton as they are in nuclear matter or other finite nuclei.

\end{abstract}

\section{Introduction}
The Triton binding energy and the related n-D scattering length $^2a$ has been
the focus of numerous investigations dating back more than 30
years.\cite{glo96} A
problem that plagued the early studies and never resolved at the time was
the relation between two-body off-shell effects and  many-body forces.
\cite{noy70,lev74}
The interest in the Triton problem and the nuclear many-body
problem in general has been revitalised by the Effective Field Theory (EFT)
approach to the study of nuclear interactions initiated by
Weinberg\cite{wei90} and
implemented by van Kolck\cite{koc01} and others. 
A guiding principle in these studies
is that data from low energy nuclear experiments should not depend on
details of the high
energy components of the underlying QCD theory of the interactions,
and that these can simply be included by 'contact'-interactions. 
A consistent derivation of nucleon interactions involves expansions ('power
counting') amounting to a
momentum cut-off usually denoted by $\Lambda$.
One important conclusion is that contrary to earlier  'dogmas'
off-shell properties of the illusory two-nucleon potential is NOT an
observable entity \cite{fur01}, an observation in line with 
work on S-matrix theory by Haag more than 50 years ago.\cite{haa58}
The  message of this result is that two-nucleon off-shell properties are
indistinguishable from many-body forces in a many-body system. Their
relative contributions ('strengths') are indistiguishable and not subject
to observation referred to as 'the equivalence theorem'. 
If regarded separately they are solely
theoretical objects that depend on the choice of the underlying QCD
Lagrangian field. Polyzou and Gl\"ockle\cite{pol90} reached the 
same conclusions using formal scattering theory.
One should of course also be aware of the fact that the potential itself is not an observable either.

Experimental information on the two-body N-N system consists of scattering
data and the bound state, the Deuteron. 
The N-N scattering data analysed in terms of  phase-shifts provides information
on the on-shell T-matrix. The Deuteron bound state provides an additional
off-shell information although this is incomplete because of the
unobservable D-state probability related to the tensor interaction. 
As stated
above other off-shell information is not observable and thus leaves
undefined the
off-shell part of the two-body
potential if constructed  solely from experimental two-body data. 
The various meson-theoretical 
NN-potentials on the market all  reproduce  on-shell
data such as phaseshifts fairly well but differ in their off-shell
predictions\cite{mac89}. 

One main purpose of finding 'the' N-N potential has been for use in the
many body theories of nuclear matter and finite nuclei e.g the triton.
Some important results of these efforts can be summarised in the 'Coester'- and
'Phillips'- lines, \footnote{relating to Brueckner theory of nuclear matter and 
the Triton respectively} that refer to binding-energies
and reflect the differences in off-shell properties of the potentials that
generate these lines. Potentials that differ in off-shell propagation differ in two-body correlations
that in turn affect the in-medium interaction. This is shown by 
eqs (\ref{domega}) - (\ref{domega2}) for nuclear matter in Section 2.

The Triton calculations are exact, being solutions of the Faddeev
equations. If this is done with some specific on-shell NN-interaction,
then the difference between the experimental and calculated  Triton binding 
energy (8.48 MeV) would
determine the 3-body contribution (or off-shell correction)
, but unique only to that
specific NN-interaction. This is the message from EFT.

The Triton probes essentially the $^1S_0$ and the $^3S_1-^3D_1$ interactions
in the two-body sector. Previous works have shown that higher angular
momentum states only contribute a few tenths of an $MeV$. To be able to 
single out the $S$-states
is very fortunate.  Firstly, because the OPE-Potential is well established for
these states. Secondly, because it can be argued that these states, in particular 
the $^1S_0$ state, for low momenta  can be well approximated by a
separable potential in the literature referred to as the 
Unitary Pole Approximation or simply UPA. 
 \footnote{This may seem
contradictory as the OPEP is local. It was however shown by Harms (see ref.
\cite{lev74}) that
contrary to expectation the OPEP is well approximated by a one-term
(rank-1) separable potential for momenta $k \leq \sim 2 fm^-1$. } 
This is a consequence of the large scattering length in this state with a
pole of the T-matrix near zero momentum, and the fact that the T-matrix
(and therefore also the potential) is separable for momenta in the vicinity
of the pole. With the on-shell T-matrix defined by the scattering data the
off-shell is then also defined. \cite{lov64}

By increasing the rank of the separable potential off-shell properties of
the $T$-matrix calculated from this potential can
be adjusted at will. This was the theme of some earlier work in which
separable potentials were used in Brueckner nuclear matter
calculations\cite{kwo95} where Deuteron data from Bonn-A,B,C potentials  and
Arndt phase-shifts for all
channels with $J\leq 5$ were used as input.
In that initial nuclear matter work, briefly reported below in Sect. 3,  only 
the lowest possible rank that was needed to
fit the data was used. Still, it was found and shown below that
the results for the $S$-states agreed
completely with the Bonn results.

In this paper we present results of Triton-calculations with separable
potentials.  
They may serve as a guide-line
for the more serious calculations that will eventually 'solve' the nuclear
many-body problem starting from QCD and/or EFT-methods rather than from the
phenomenological approach used here.

The method of inverse scattering with separable potentials is 
briefly reviewed in Sect. 2.
Sect. 3 shows some results of earlier Brueckner calculations for
nuclear matter that are relevant for the present work. 
Sect. 4 presents results of Triton and $n-D$ calculations,
while Sect. 5 presents a summary and a discussion of results.

\section{Separable Interaction from inverse scattering}
The methods used here to calculate a separable interaction by inverse
scattering were
used in several previous papers relating to Brueckner and Green's function
calculations of nuclear and neutron matter as well as the Unitary problem. 
\cite{kwo95,hsk04,hskm07,hsk07,hsk08,hsk108}
The input in the calculations are 
phase-shifts and Deuteron data.
A potential is derived for each two-body state from these data. 
A rank-1 separable potential is sufficient for the on-shell fit if all
phases have the same sign.
If the phase-shift
changes sign such as in the  $^1S_0$ case, a  rank-2 potential
is the minimal requirement. 
An increased  rank  allows for
off-shell changes, keeping the on-shell intact.
The Deuteron data contains some off-shell information requiring
a rank-4 potential.  For briefness
only the rank one formalism is shown below. (See ref \cite{kwo95} for more
details).
 With a rank one attractive potential given by
\begin{equation}
V(k,p)=-v(k)v(p)
\label{V}
\end{equation}
inverse scattering gives
(e.g. ref \cite{kwo95,tab69})

\begin{equation}
v^{2}(k)= \frac{(4\pi)^{2}}{k}sin \delta (k)|D(k^{2})|
\label{v2}
\end{equation}
where
\begin{equation}
D(k^{2})=exp\left[\frac{2}{\pi}{\cal P}\int_{0}^{\Lambda}
\frac{k'\delta(k')}{k^{2}-k'^{2}}dk' \right]
\label{D}
\end{equation}
where ${\cal P}$ denotes the principal value  and
$\delta(k)$ is the input phaseshift. $\Lambda$ provides a cut-off
(renormalisation) in
momentum-space with $\delta(k)=0$ for $k>\Lambda$.
The effect of the  cut-off will be exploited below.

With the interaction ${\bf V}$=$V(k,p)$ and a kernel 
${\bf G}$=$G({\bf k},{\bf P},\omega)$, where ${\bf P}$ is the center-of-mass
momentum one
can define an 'effective' interaction ${\bf K}_G$ by a Lippman-Schwinger equation:
\begin{equation}
{\bf K}_G={\bf V}+{\bf V}{\bf G}{\bf K}_G
 \label{KG}
 \end{equation}
With the rank-1 separable interaction given by eq. (\ref{V}) , the solution of this
equation is simply
\begin{equation}
 {\bf K}_G({\bf k},{\bf p},{\bf P},\omega)=-\frac{V(k,p)}{{\cal D}_{G}
 ({\bf P},\omega)}
 \label{G}
 \end{equation}
 where the potential is here assumed to be attractive and where
 \begin{equation}
 {\cal D}_{G}({\bf P},\omega)=1+\frac{1}{(2\pi)^{3}}\int_{0}^{\Lambda}
 v^{2}(k)G({\bf k},{\bf P},\omega)
 k^{2}dk
 \label{I_G}
 \end{equation}
and ${\bf K}_G$ is  separable in functions of $k,p$ and a function 
of $(P,\omega)$ .

 Three  separate kernels and the associated matrices will be considered here:\\
 \\
 1) $G=(\omega-k^2+i\eta)^{-1}$\\
 Then \\
 ${\bf K}_G\equiv$  ${\bf T}({\bf k},{\bf p},\omega)$; the
 scattering-matrix\\
 with\\
 ${\bf T}(k,k,k^2)=e^{i\delta(k)}sin\delta(k)/k$\\
 \\
 2) $G={\cal{P}}(\omega-k^2)^{-1}$\\
 where $\cal P$ implies the principal value.\\
 Then \\
 ${\bf K}_G\equiv$  ${\bf R}({\bf k},{\bf p},\omega)$; the
 reactance-matrix \\
 with\\
 ${\bf R}(k,k,k^2)=tan\delta(k)/k$\\
 \\
 In the two cases above the 'effective' interactions are independent of ${\bf
 P}$.\\
 3) $G=Q({\bf k},{\bf P})(\omega-e({\bf k},{\bf P}))^{-1}$\\
 where $Q$ is the Pauli blocking operator and the single particle energy 
 $e({\bf k},{\bf P})$
 includes a self-consistent mean field (self-energy).\\
 Then \\
 ${\bf K}_G\equiv$  ${\bf K}({\bf k},{\bf p},{\bf P},\omega)$; 
 the Brueckner reaction-matrix, in the literature sometimes denoted by
 ${\bf G}$. \\
 \\

With the potential derived from inverse scattering it follows that
the diagonal on-shell elements of the reactance matrices (${\bf R}(k,k,k^2)$)
are by definition independent of $\Lambda$ (for $k<\Lambda$ of course).
Effective in-medium  interactions, e.g. the Brueckner $K$- (or $G$-) 
matrices differ from
these $R$ due to the Pauli-operator and the propagator self-energy in the
definition of $K$, bringing it off-shell.

Off-shell effects are of particular interest in many-body problems.
To illustrate the origin of this effect let us consider the Reaction
matrix ${\bf K}$ defined above, and estimate the effect of a change in the
selfconsistent propagator i.e. the dependence on $\omega$.\footnote{In
Brueckner calculations the $\omega$ is a single particle energy $e({\bf
k},{\bf P})$ that includes the mean field.} One finds
 \begin{equation}
 {\bf K}({\bf k},{\bf p},{\bf P},\omega')={\bf K}({\bf k},{\bf p},{\bf P},\omega)+
\int d{\bf k'} {\bf K}({\bf k},{\bf k'},{\bf P},\omega)
\left(\frac{Q({\bf k'},{\bf P})}{\omega-e({\bf k'},{\bf P})}
 -\frac{Q({\bf k'},{\bf P})}{\omega'-e({\bf k'},{\bf P}) }
 \right){\bf K}({\bf k'},{\bf p},{\bf P},\omega')
 \label{domega}
 \end{equation}

Using 
 \begin{equation}
 {\bf \Psi}({\bf k},{\bf p},{\bf P},\omega)={\bf \Phi}({\bf k},{\bf p})+
\int d{\bf k'} {\bf \Phi}({\bf k},{\bf k'})\frac{Q({\bf
k'},{\bf P})}{\omega-k'^2}
 {\bf K}({\bf k'},{\bf
 p},{\bf P},\omega)
 \label{psi}
 \end{equation}
where $\Psi$ and $\Phi$ are the in-medium two-body correlated and uncorrelated
wave-functions respectively one finds 
 \begin{equation}
 {\bf K}({\bf k},{\bf p},{\bf P},\omega')={\bf K}({\bf k},{\bf p},{\bf P},\omega)+
 I_w(\omega-\omega')
 \label{domega1}
 \end{equation}
 where $I_w$ is a "wound"-integral defined by.
 \begin{equation}
 I_w=\int ({\bf \Psi}_{k,P}(r)-\Phi_k(r))^2d{\bf r}
 \label{domega2}
 \end{equation}

 With $\omega-\omega'=$ potential-energy (self-energy-insertion),  
 the last term in eq (\ref{domega1}) is referred to as a dispersion correction.

The off-shell-effect that generates this correction is important for
saturation and nuclear binding in general because the self-energy 
is density-dependent. It is important to note that  by eq. (\ref{domega2})
this off-shell effect is directly related to (un-observable)  correlations. 
It is explicitly a three-body
effect but not to be understood as a three-body force because it \it "is built up
out of two-nucleon interactions"\rm. \footnote{The quote is from ref.
\cite{wei90}}  The effect that the  momentum-cut-offs has on the two-body 
correlations was  a subject discussed in
ref.\cite{hskm07}. It was there already stressed  that the relation between 
off-shell and correlation effects
plays an important role in nuclear many-body physics and is intimately
related to three-body correlations. 
A major purpose of this work is to show that these correlations play an important role 
also in the Triton-case.
The word 'realistic' is used frequently to emphasize the quality of a
specific potential, mostly referring to the fits to phase-shifts. In that
sense our separable potentials are realistic as they fit phase-shifts
exactly by construction. An added virtue is the agreement with the
accepted realistic Bonn-potentials in the Nuclear Matter results shown
below.

\section{Nuclear Matter}
\begin{figure}
\centerline{
\psfig{figure=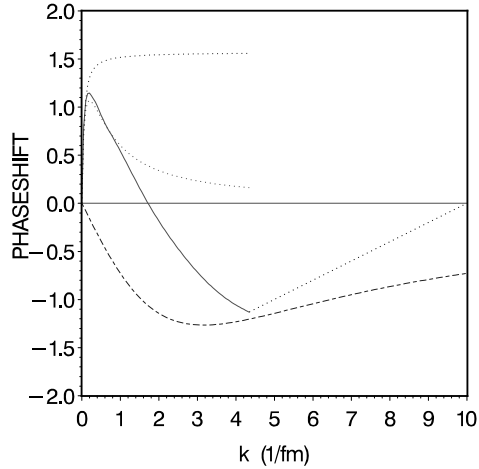,width=7cm,angle=0}
}
\vspace{.0in}
\caption{The Arndt $^1S_0$ phaseshifts are shown by the solid line. Its
continuation by a dotted line to $k=10 fm^{-1}$ shows phases used in
some calculations as referred to in the text.
Upper dotted curves show
scattering length and effective range approximations. 
The broken curve at
the bottom shows the repulsive phaseshifts defined in eq. (\ref{hrc}) with
$c=0.1$ and $r_c=0.8$ which were used in some of the calculations on the Triton
shown below in Sect.4. 
}
\label{trit1}
\end{figure}

Some earlier reports  on nuclear matter calculations with separable
potentials\cite{kwo95,hsk00}  that are relevant for the present
Triton calculations 
are summarised in this section.
The scattering data that was 
used as input in the construction of the potentials were 
the Arndt 
phase-shifts\cite{arn87} (Fig. \ref{trit1} shows $^1S_0$ phases.) 
 for all channels with $J\leq 5$ . 
Other input were Deuteron properties as defined by either of the
Bonn-A,B or C potentials. 
The original theme of this early work was to explore
off-shell dependence of the in-medium interactions (easily achieved by
increasing the rank) and its effect on nuclear properties.
Considered as the first step only
 the lowest possible ranks needed to
 fit the data were used.  Even so it was found that the 
 results of our Brueckner calculations for the $S$-states agreed very well
with those of the  Bonn Brueckner results.
Table I shows  our contributions to the energy/particle from
the $^1S_0$ and $^3S_1$ states
compared with the BONN-B results\cite{mac89} . 
The fermimomentum $k_f$ is in units of $fm^{-1}$ and the energies
in $MeV$ per particle.
(See  ref.  \cite{kwo95} for  details.)
\newpage
\begin{tabbing}
\hspace{2.0 in} TABLE I \\
\\
================================================\\
\hspace{1.2in} $^1S_0$ \hspace{2.2in} $^3S_1$\\
-------------------------------------------------------------------
----------------------------------------------\\
$k_{f} $ \hspace{0.2in}\=\hspace{.0in} BONN-B \hspace{.5in}\=SEPARABLE 
\hspace{.5in} \=BONN-B \hspace{.5in}\=SEPARABLE\\
================================================\\
1.35 \> \hspace{0.05in}$-16.66$ \>\hspace{0.05in}$-16.57$ \>
\hspace{0.05in}$-21.34$ \>\hspace{0.05in}$-21.33$\\

1.60 \> \hspace{0.05in}$-22.62$\> \hspace{0.05in}$-22.76$ \>
\hspace{0.05in}$-26.59$ \>\hspace{0.05in}$-26.27$\\

1.90 \> \hspace{0.05in}$-28.72$\>\hspace{0.05in}$-29.84$\>
\hspace{0.05in}$-31.36$ \>\hspace{0.05in}$-31.45$\\
================================================ \\
\end{tabbing}

The (almost) complete agreement between  our separable 
and the BONN results may seem fortuitious.
Although the respective potentials  are (almost) phase-shift equivalent,
any off-shell agreement is
not guaranteed. The methods of constructing the potentials 
are indeed widely different and while the BONN are local  the separable are
non-local. 
The agreement 
for the $^1S_0$ state may however also be a direct consequence of that the 
$^1S_0$-potential \it is \rm separable. (See also Sect. 1).  
In the $^3S_1$ case the argument is
somewhat different. Both fit the same Deuteron-data 
 which implies a fit not only to scattering data but also to
an off-shell energy, the binding energy of the Deuteron. \footnote{For a
detailed discussion see ref. \cite{kwo95}}
This fit also includes the Deuteron wave functions  (in the cases shown
the Bonn-B). At this point it should be observed that these are related 
to the un-observable D-state probability $P_D$. Results of Brueckner
(and finite nucleus) calculations  show  a dependence on $P_D$. 
According to EFT, a correct calculation must then include 
counter-terms that eradicate this dependence. 

The agreements in Table I are consistent with
the agreements with the BONN half-shell
reactance matrices for these states as shown in ref. \cite{kwo95} (See
figs 4 and 6 in this reference.) 
\footnote{The reason for showing Reactance rather than e.g. Brueckner
Reaction matrix half-shell is that the latter as shown in Sect. 2 
depend on 4 rather than 2 variables.}
For the $SD$ and $DD$ matrices these agreements were however
not good and correspondingly there was a $0.55 MeV$ difference in the $^3D_1$
contribution to the binding. The sources of these agreements and 
discrepancies were explained in more detail in ref.\cite{kwo95}.
Less good agreements were also found in states for which,  unlike the
$S$-states their corresponding $T$-matrix does not have a pole close to
zero energy.
Noticeable was in particular the disagreement in the $^3P_1$ state with
our binding $1.48 MeV$ more repulsive than the BONN. 
This was consistent with more repulsion
in off-shell reactance matrix elements shown in Fig. 5 of ref.\cite{kwo95}
A rank one potential was used for this state as this was sufficient for
fit of the phase-shifts in this case. With an increase to rank-2 the
off-shell repulsion was corrected and simultaneously the binding from this
state.\cite{hsk00} 
 It should also be observed that for many of the high
angular momentum states with small
phase-shifts the \it phase-shift approximation\rm \cite{ries56}  
is good, making  potentials for these states less needed.

The effect of varying  high-energy phase-shifts (beyond
those determined experimentally ) was investigated already 
in ref. \cite{kwo95}. Although the potentials would change with such
variations, on-shell properties defined by the known phase-shifts would
of course remain the same. In contrast, half-shell Reactance matrices  did
as expected change, but less so than the potentials, especially for the
low momenta relevant for nuclear structure calculations. The high-energy
phases of choice were therefore the straight line extension of the Arndt
phases shown in Fig. \ref{trit1}. Larger changes
simulating increased short ranged repulsions are used below (Sect. 4.)
in the Triton
calculations affecting also nuclear matter results.

\begin{figure}
\centerline{
\psfig{figure=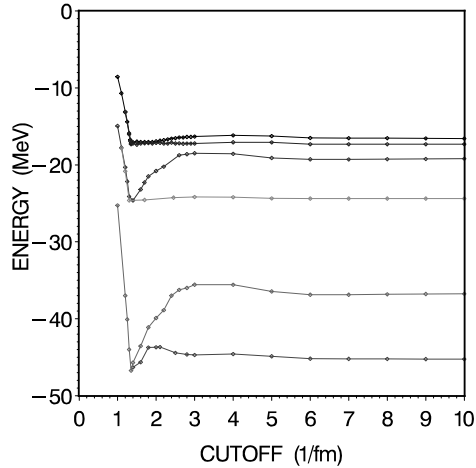,width=7cm,angle=0}
}
\vspace{.0in}
\caption{
Binding energies as a function of $\Lambda$ are shown here.
There are three sets of curves, the uppermost shows the contribution to
the potential energy per particle  from the
$^{1}S_{0}$ state the middle from the $^{3}S_{1}$ and the bottom
includes all (21) states. 
Shown are also the effects of a self-energy insertion in
the nucleon propagators (dispersion-correction) 
i.e. a three-body term. See text for further details and discussion of
these results.
}
\label{trit2}
\end{figure}

The effect of cutoffs $\Lambda$, with phaseshifts 
equal to zero for $k>\Lambda$ was investigated in some later reports.
\cite{hsk04,hskm07}. Some results from these that are relevant for comparison
with the Triton calculations are shown in Fig. \ref{trit2}. (From 
ref. \cite{hskm07}.) It shows results of Brueckner calculations of
potential energy contributions in
nuclear matter at saturation density. 
In each of the three sets of curves there is a lower and an upper curve.
In the propagator $G$ that defines the Brueckner $K$-matrix the mean field
(self-energy) is included in the upper curve, but not in the
lower. 
This off-shell effect generated by the mean field is seen to be repulsive  but 
starts to decrease when the
cutoff $\Lambda$ decreases below $\Lambda \sim 3.0 fm^{-1}$ and
approaches zero as $\Lambda \rightarrow k_{F}=1.35 fm^{-1}$. For
$\Lambda<k_f$ the range of momenta that contribute to the binding will be
too small and hence the sharp
decrease in binding. The off-shell effect is seen to be much
smaller for the singlet than it is for the triplet $S$ state. 
Eq. (\ref{domega1}) above shows an important relation 
between the two-body correlation and these off-shell effects.
From this relation one would conclude that the correlations are much
larger (the wound-integral much larger) in  case of the triplet than in the
singlet.
Tensor-correlations are absent in the $^1S_0$-state. In this state the 
correlation is mainly due to the short ranged repulsion and is smaller. 
Figs (5-8) in 
ref.\cite{hskm07} shows a difference between the  correlated wave-functions 
for the two states
in accordance with the  above. Further evidence is the near absence of
correlations shown in ref.\cite{hskm07} Figs (9,10) when $\Lambda=2 fm^{-1}$ .

The effect of the self-energy insertion in the Brueckner propagators can be
interpreted in two different ways. i)As a change in the interaction
(correlation) between two nucleons i and j due to the presence of a third k or 
ii) as the reduced interaction of nucleon k with i because i is 
correlated with (partly excited by) nucleon j reducing  occupation $n_i$. 
The first interpretation invokes a three-body \it force \rm because it
explicitly pictures the
modification of the interaction between two nucleons due to the
presence of a third. 
It is however built up out of two-nucleon interactions and is not
intrinsically a three-body force, and is in nuclear many-body theory
(mostly)
referred to as a three-body term. (See als end of Sect. 2) It is a consequence of two-body
correlations that are (nearly) independent of the medium. There is some
medium-dependence here but already included by the $Q$-operator and the
mean-field itself.
A three-body \it force \rm on the other hand would be due to a change
of these correlations due to a presence of a third in the medium, e.g. by a
polarisation of the mesonic fields. 

Some further conclusions regarding off-shell (three-body) effects can be
deduced from Fig. \ref{trit2} as follows.
For each value of $\Lambda$ in Fig. \ref{trit2} a new  potential is 
calculated for each state. Consider first the upper lines in each of the
three sets. Although the potentials change, the figure shows that all binding 
energies are essentially constant for  $\Lambda>\sim 3 fm^{-1}$.
Consider the Brueckner $K$-matrix from which the curves are calculated 
to be the effective two-body interaction that includes the three-body term
discussed above.
Invoking EFT, discussed above, a proper renormalisation of two- and three-body \it
forces \rm
should leave the calculated energy independent of $\Lambda$. 
This implies that contributions from 3-body forces  are constant
for $\Lambda>\sim 3 fm^{-1}$, because the two-body alone is constant there
but also that the 3-body forces would have to become increasingly repulsive 
for $\Lambda<\sim 3 fm^{-1}$. But a change in 2-body off-shell repulsion
can also have the same effect. It can be done by adding a contact force.
This will be done in the 3-body calculations below.

The difference between the two curves, the lower and the upper in each of
the three sets is a consquence of the relation given by eq. (\ref{domega1}))
the dispersion-correction.  It contributes a
repulsive medium(density)-dependent component to the effective interaction.\cite{mos60}
In Brueckner calculations this is a major contribution to the saturation of 
nuclear matter. It likewise contributes an important component (often referred to as a "rearrangement" potential) 
to the selfconsistent Brueckner Hartree-Fock potential in
finite nucleus calculations. Without it the finite nucleus collapses. 
It justifies (in part) the density dependent part of the Skyrme interactions. 

The Brueckner theory does of course not give an exact solution of the many 
body problem while the Faddeev equation does give an exact solution of 
the 3-body problem for a given set of potentials. 
It therefore seems of interest to find the results of using  our  potentials in
Triton and n-D calculations.

\section{Three-body Calculations}
The three-body problem in nuclear physics has been a subject of interest
for a long time. $^3H$ and $^3He$ calculations showed a dependence on
two-body off-shell properties. 
Three-body forces were introduced (e.g. ref.\cite{coo79}  to seek 
improved fits to experimental data.\cite{wit03}
The three-body 
system also provides a good testing ground for EFT methods.(e.g.
refs\cite{bed00,afn04,she09})

The formalism necessary for calculating the Triton  energy $E_T$ and the
$n-D$ scattering length $^2a$ using the Faddeev equations \cite{fad60}
is well documented. Calculations are
substantially simplified using separable rather than local
potentials with numerous reported results. Separable
parametrizations of the Paris potential has for example also been used in
three-nucleon calculations.
\cite{hai86}
Only the simplest version of the Faddeev equation(s)
for a spin-independent rank-1 separable interaction acting 
only in the two-body $S$-state is shown here:
\begin{equation}
\chi(q)=\frac{2}{{\cal D}(E_T-\frac{3}{4}q^2)}\int \frac{v(|{\bf k}+\frac{1}{2}{\bf
q}|)v(|{\bf q}+\frac{1}{2}{\bf k}|)}{q^2+{\bf q}\cdot {\bf
k}+k^2-E_T}\chi(q)d{\bf k}
\label{faddeev}
\end{equation}
with (cf eq. (\ref{I_G})
\begin{equation}
{\cal D}(s)=1+\frac{1}{(2\pi)^{3}}\int_{0}^{\Lambda}
v^{2}(k)(s-k^2)^{-1} k^{2}dk
\label{DG}
\end{equation}

The notations are the standard: $E_T$ is the Triton energy, ${\bf k}$ is
the relative momentum of two particles and ${\bf q}$ is the momentum of the
third. This equation is conveniently solved for $E_T$ using the Malfliet-Tjon
method \cite{mal69}.

\begin{figure}
\centerline{
\psfig{figure=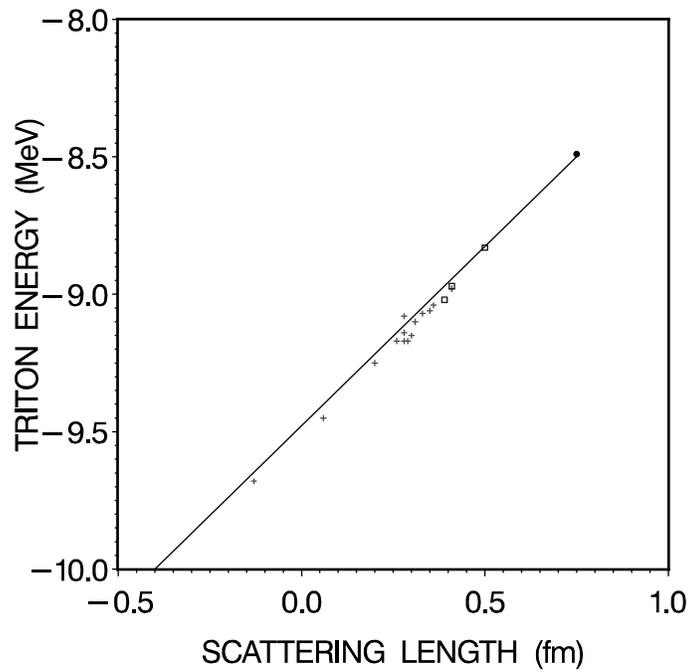,width=10cm,angle=0}
}
\vspace{.0in}
\caption{
The solid line is the Phillips line defined by eq. (\ref{phil}). The
crosses are results with the separable potentials from inverse scattering
using the Arndt phase-shifts and with the Bonn-B Deuteron parameters while
the squares are with
Bonn-C.  The dot at the top of the Phillips line is the experimental point. 
 See also Fig. \ref{trit6}. 
}
\label{trit3}
\end{figure}

\begin{figure}
\centerline{
\psfig{figure=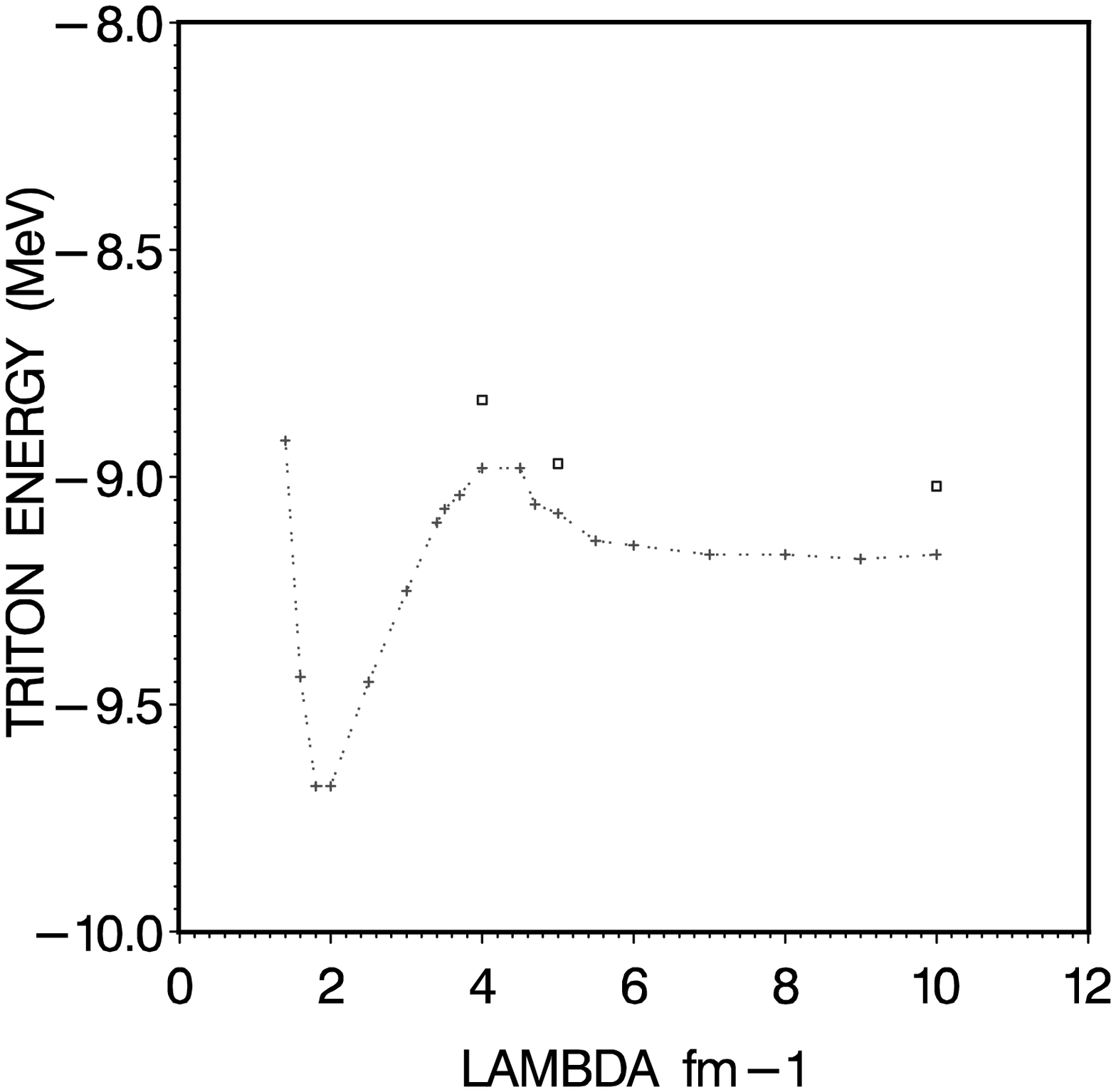,width=10cm,angle=0}
}
\vspace{.0in}
\caption{
Shown is the Triton energy as a function of cutoff.
The crosses joined by broken lines are with Bonn-B Deuteron parameters
while the squares are with Bonn-C.
}
\label{trit4}
\end{figure}

\begin{figure}
\centerline{
\psfig{figure=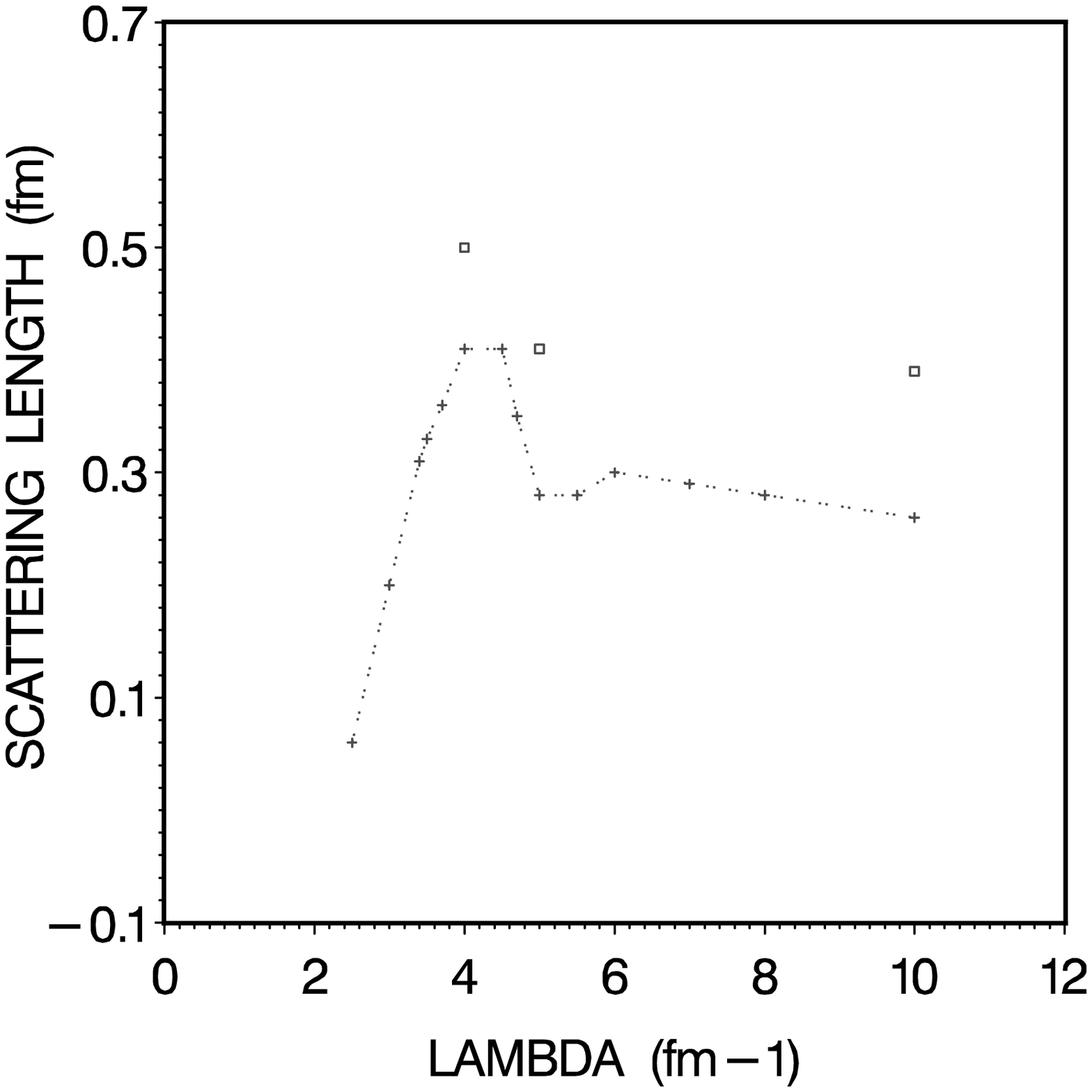,width=10cm,angle=0}
}
\vspace{.0in}
\caption{
Shown is the $n-D$ scattering length $^2a$  as a function of cutoff.
The crosses joined by broken lines are with Bonn-B Deuteron parameters
while the squares are with Bonn-C.
}
\label{trit5}
\end{figure}

The equations relevant for the present calculations with spin-dependent
and higher rank potentials are, for example, found in refs 
\cite{fud68,sta69}. 
The paper by Dabrowski et al\cite{dab71} is also
exceptionally helpful with detailed presentation of the formalism in
particular regarding $^2a$. 
Fig. \ref{trit3} shows our results for $E_T$ vs $^2a$ for  our separable
potentials for the different values of $\Lambda$ shown in Figs \ref{trit4}
and \ref{trit5}. 
Early work with separable
potentials were found to lie along a "Phillips" line\cite{phi68}. Harms\cite{har72}
fits these data by   (see ref.\cite{lev74}).
\begin{equation}
^2a=0.75(E_T+8.5)+0.75 fm
\label{phil}
\end{equation}
where $E_T$ is in $MeV$. Our results are seen to agree with these early
results that are also supported by the EFT method.\cite{bed00}
Fig. \ref{trit4} shows the Triton energy as a function of cutoff.
Figs \ref{trit4} and \ref{trit5} show that both $E_T$ and $^2a$ are fairly
constant for $\Lambda\geq 5 fm^{-1}$ with maxima at $\sim 4 fm^{-1}$i
followed by a minimum at $\sim 2fm^{-1}$(not shown in Fig. \ref{trit5}). The
decrease in energy for small $\Lambda$ is similar to that for nuclear
matter as shown in Fig. \ref{trit2} at approximately the same $\Lambda$ but
the
Triton results differ from those of nuclear matter in the sharp increase at
$\Lambda \sim 4 fm^{-1}$.

Nogga et al\cite{nog04} show $V_{low-k}$ results of $E_T$ as a function 
of $\Lambda$ for AV18 and CD Bonn potentials. Their curves also show minima
although at a slightly smaller value of $\Lambda$, $1.6 fm^{-1}$ compared
to ours at $2 fm^{-1}$. They do not show the increase at around $4 fm^{-1}$.
More significant is that separable potentials in general show appreciably
over-binding of the Triton as opposed to local potentials that show
under-binding. This has of course been known for  a long time, but this
difference has never been well understood. 
As shown above in Table I, there is
almost perfect agreement for nuclear matter in the $S$-states so why is
there a
large difference in Triton energy where $S$ states dominate. The answer
must be simply that the two cases, nuclear matter and the Triton explore
different parts of the interactions. 
One can however conclude that
the dependence on cutoffs $\Lambda$ for the Triton, reminiscent of that for
nuclear matter is (in our case) understood as a reflection of the change in
off-shell properties of the potentials as a function of $\Lambda$.

Fig. \ref{trit5} shows the $^2a$ scattering length as a function of cutoff.
It shows a $\Lambda$ dependence very much like in Fig. \ref{trit4} for the
energy. This is consistent with the Phillips line in
Fig.\ref{trit3}.

Our results show an over-binding of the Triton.
It is of interest to find if and what modifications of the 
interaction can be made to decrease the binding and if this would or would
not mean a shift off the Phillips line. It seems apparent that the strength
of the in-medium interaction   would have to be decreased but 
can this be done without changing the low momentum on-shell interaction
that is fixed experimentally?
In this equation the form-factors $v(k)$ in the numerator are only needed
for the \it low \rm momenta $k$ of nucleons in the Triton.  They are fixed by the
input data. The only possible freedom of change  is in the denominator 
${\cal D}_G$ that depends on the medium and involves an integration over
\it high \rm momenta. The form-factors $v(k)$ at these high
momenta relate to the un-known short ranged part of the potential that
therefore leaves room for a parametrisation.
One can argue that the potential defined by eq. (\ref{v2}) is a functional of
$\delta(k)$ i.e. a function of the phases at all momenta.  As a consequence
it is not possible to independently vary high and low components of the
form-factor $v(k)$ by a similar change in phases. 
Rather than the potential it is the high and low momentum components of the
in-medium interaction that should be discussed but it was found in ref.
\cite{kwo95} that high energy phases only weakly affect low-energy parts of
the reactance-matrix which is a fair approximation of the effective interaction.
The potential on the other hand was in fact shown to be affected for all
momenta.

Our discussion above together
with eq. (\ref{domega1}) implies that a short-ranged repulsive potential
increases two-body correlations and induces a repulsive term in the effective
in-medium interaction. And it is easily incorporated in the potential without
destroying the fit to low-energy data.

Incorporation of a short-ranged repulsive potential is done by using a 
rank-2 separable potential as follows (also used in earlier works,
e.g.\cite{dab71,fie70}):
\begin{equation}
V(k,p)=h(k)h(p)-g(k)g(p)
\label{hhgg}
\end{equation}
with a repulsive and an attractive part respectively. 
The $h(k)$ form-factor was
determined by an inverse scattering method using repulsive phase-shifts
of the following form 
\footnote{A constant repulsion, a contact force, is expected to have the
same effect, but the form chosen seemed more practical .}
\begin{equation}
\delta_s(k)=kr_c/(1.+c*k^2)
\label{hrc}
\end{equation}
where $r_c$ would be related to the range of the assumed
repulsion.
The constant $c=0.1$ was chosen to limit $\delta(10 fm^{-1})$ to $<\pi/2$,
$10 fm^{-1}$ being the cutoff used in these particular calculations.
With $h(k)$ given, the form-factor $g(k)$ was  determined by inverse
scattering to reproduce the experimental phase-shifts known only for
$k\leq 4.35 fm^{-1}$ but extended to $10 fm^{-1}$   by our
parametrisation of the repulsive part as shown in Fig.\ref{trit1}. 
The inverse scattering method used in ref \cite{kwo95} for the $^1S_0$-state 
was that of Bolsterli and McKenzie\cite{bol65} while with the present 
rank-2 potential the
method of Fuda was used instead. \cite{fud70}   In the previous work the
Fuda method was
only used for
the $^3S_1$ potential. 

\begin{figure}
\centerline{
\psfig{figure=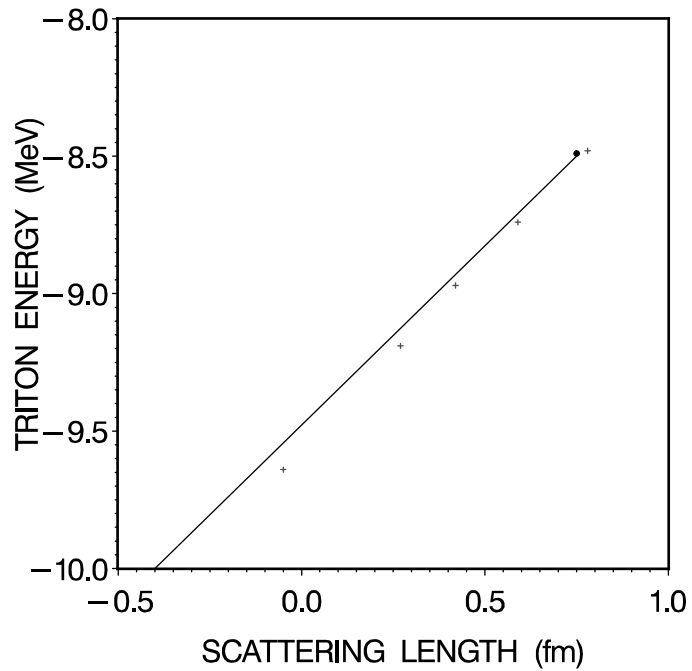,width=10cm,angle=0}
}
\vspace{.0in}
\caption{
Similar to Fig.\ref{trit3} but with the $^1S_0$ phase-shifts modified as
explained
in the text. The deuteron paramaters are those of Bonn-B. The calculated
point close to the experimental (the dot) is obtained by a fit adjusting
high energy phase-shifts to $r_c=0.8 fm^{-1}$ while $c=0.1$ (see text).
}
\label{trit6}
\end{figure}

\begin{figure}
\centerline{
\psfig{figure=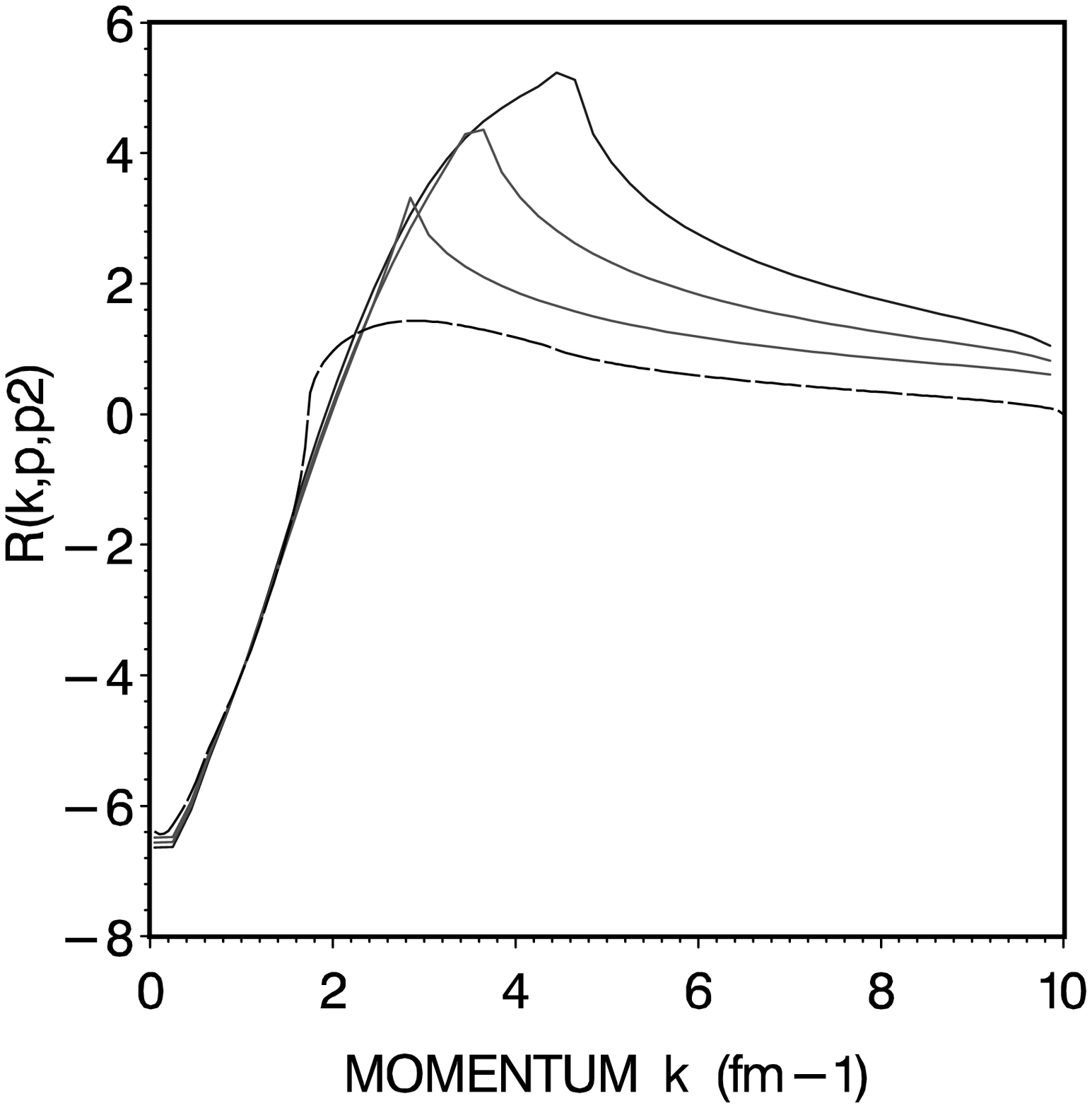,width=7cm,angle=0}
}
\vspace{.0in}
\caption{
Solid lines show
half-shell reactance matrix elements for three different values of
short-range repulsive parameter $r_c$ . Parameter ranges from $0.4 fm$
(bottom solid curve) to $0.8 fm$ (top solid curve) and  momentum $p=1 fm^{-1}$.
The broken (lowest) curve is without any short-ranged repulsion i.e. with the Arndt
phases extrapolated to $\delta(10 fm^{-1})=.0$ as shown by the dotted curve in
Fig. \ref{trit1}.
}
\label{trit7}
\end{figure}

Results of $E_T$ vs $^2a$ are shown by the crosses in Fig. \ref{trit6}
Like the earlier results shown in Fig. \ref{trit3} they lie along a line 
although slightly off from the the Phillips-line defined by
eq.(\ref{phil}). The three crosses are from bottom to top obtained with
$r_c=0., 0.5, 0.6, 0.7$ and $0.8 fm$ repectively with the latter being very
close to the experimental point shown by the dot. 
This upward movement along the Phillips-line with increased short-ranged
repulsion   has been
shown earlier. \cite{dab71} Using the arguments above an increase in short
ranged high-momentum repulsion results in  an
increased repulsion in the diagonal elements of the low-momentum 
\it effective  \rm interaction because of the increased 2-body
correlations, that are due to 
a change in un-observable high-energy off-diagonal elements.
This interpretation is substantiated by Fig. \ref{trit7} showing the half-shell
reactance matrix, $R(k,p,p^2)$, for $r_c=0.4,0.6$ and $ 0.8 fm$. The
lowest, broken curve is with the 'original' phases shown by the dotted
curve in Fig. \ref{trit1}. 
In agreement with comments
above,  $R$ is  independent of $r_c$ for $k<\sim3$ but shows significance 
dependence on $r_c$ for the larger momenta. 

The result of this part of our investigation may be interpreted as
showing the role of the three-body force. The two-body potentials in Fig.
\ref{trit7} are
different at each
point along the Phillips-line but are all phase-shift equivalent. 
They all fit the same
on-shell data for $k\leq \Lambda$. They differ by yielding different in
medium (non-observable) off-shell interactions. Alternatively if not
modifying the 2-body off-shell one can instead add a three-body force to
reproduce  the Triton data. 
Of course, EFT indicates that the two statements are equivalent.

The rather drastic change in off-shell NN effective interactions has a likewise
effect on Brueckner nuclear matter calculations. The contribution to
binding in the $^1S_0$ and $^3S_1-^3D_1$ states at $k_f=1.35 fm^{-1}$ was in
the earlier calculations\cite{kwo95} $35.77 MeV/A$. With the repulsive
core calculations this contribution to binding now varies between $36.22$
for $r_c=0 $ to $31.79$ for $r_c=.8 fm$. (These particular results 
are all obtained
with $\Lambda=10 fm^{-1}$.) So  the trend is the same for 
the Triton and the Nuclear Matter calculations, but while the Triton
binding can be reproduced it leaves Nuclear matter even more under-bound.
Remember however that the Triton-calculation is exact while the Brueckner
is not. 

(The earlier result $35.77 MeV/A$ and the recent $36.22 MeV/A$ 
are both for $r_c=0$.
The (small) difference is because of the different inverse scattering
methods as described above that may result in some  off-shell  difference.
The two methods should give on-shell equivalent potentials.)

The  short ranged repulsion represented 
by the formfactor $h(k)$ is  believed to be related to the 2-body 
contact term in EFT.\cite{koc01} 
In the renormalisation process, decreasing 
$\Lambda$ as in the calculations leading to the results shown in Figs
\ref{trit3} to \ref{trit5}
the off-shell (three-body) contributions are also cut. Without
the contact term the half-shell reactance matrix $R(k,p,p^2)=0$ for
$k>\Lambda$. So while the on-shell parts of $R$ are left essentially untouched by
the renormalisation, the off-shell parts are decreased. That is the source
of the triton energy $E_T$ in Fig. \ref{trit4} not being independent of
cut-off as required by a proper renormalisation. A proper renormalisation
procedure has to include three-(many-) body forces or (according to
ref.\cite{fur01})  equivalently, maintain off-shell parts of the
interaction.

\begin{figure}
\centerline{
\psfig{figure=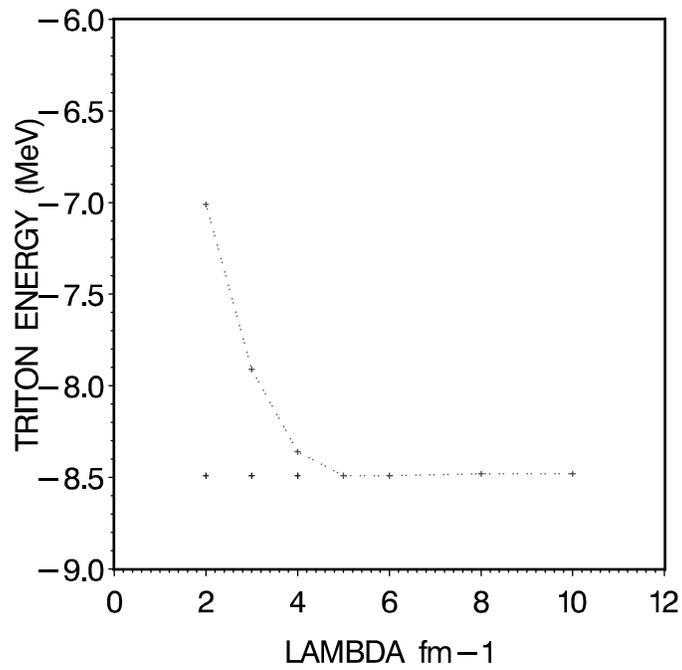,width=10cm,angle=0}
}
\vspace{.0in}
\caption{
The crosses joined by a broken line shows Triton energy as a function of
$\Lambda$ with $c=0.1$ and $r_c=0.8$. The three separate crosses show
results after adjusting $r_c$ as explained in the text. 
}
\label{trit9}
\end{figure}

This is illustrated phenomenologically by keeping the contact force 
$h(k)$ in eq. (\ref{hhgg})  intact,  but cutting the attractive
force $g(k)$ such that $g(k)=h(k)$ for $k>\Lambda$. 
Calculations were then made for a few values of $\Lambda$  and for each
value of $\Lambda$ the repulsion, the parameter $r_c$ in the contact force
$h^2(k)$,  was adjusted 
to maintain the experimental value  $E_T\sim -8.5 MeV$ and simultaneously also
 $^2a=0.78 fm$ as shown by the point close to the experimental point in Fig.
 \ref{trit6}. The adjustments were made keeping $c$ in eq. (\ref{hrc})
constant at $c=0.1$ and resulted in  $r_c=0.49,0.68$ and $ 0.78 fm$ 
for $\Lambda=2,3$ and $4 fm^{-1}$ respectively.
Thus the smaller the cut-off the weaker also the required repulsive  contact force.
If instead $c$ and $r_c$ are both kept constant while cutting the attractive force,
the Triton energy will increase and so will $^2a$ while the two move along
the Phillips line. Fig. \ref{trit9} shows the Triton energy as a function
of $\Lambda$ in this case together with the three crosses when $r_c$ is
adjusted to values just shown above. 

\begin{figure}
\centerline{
\psfig{figure=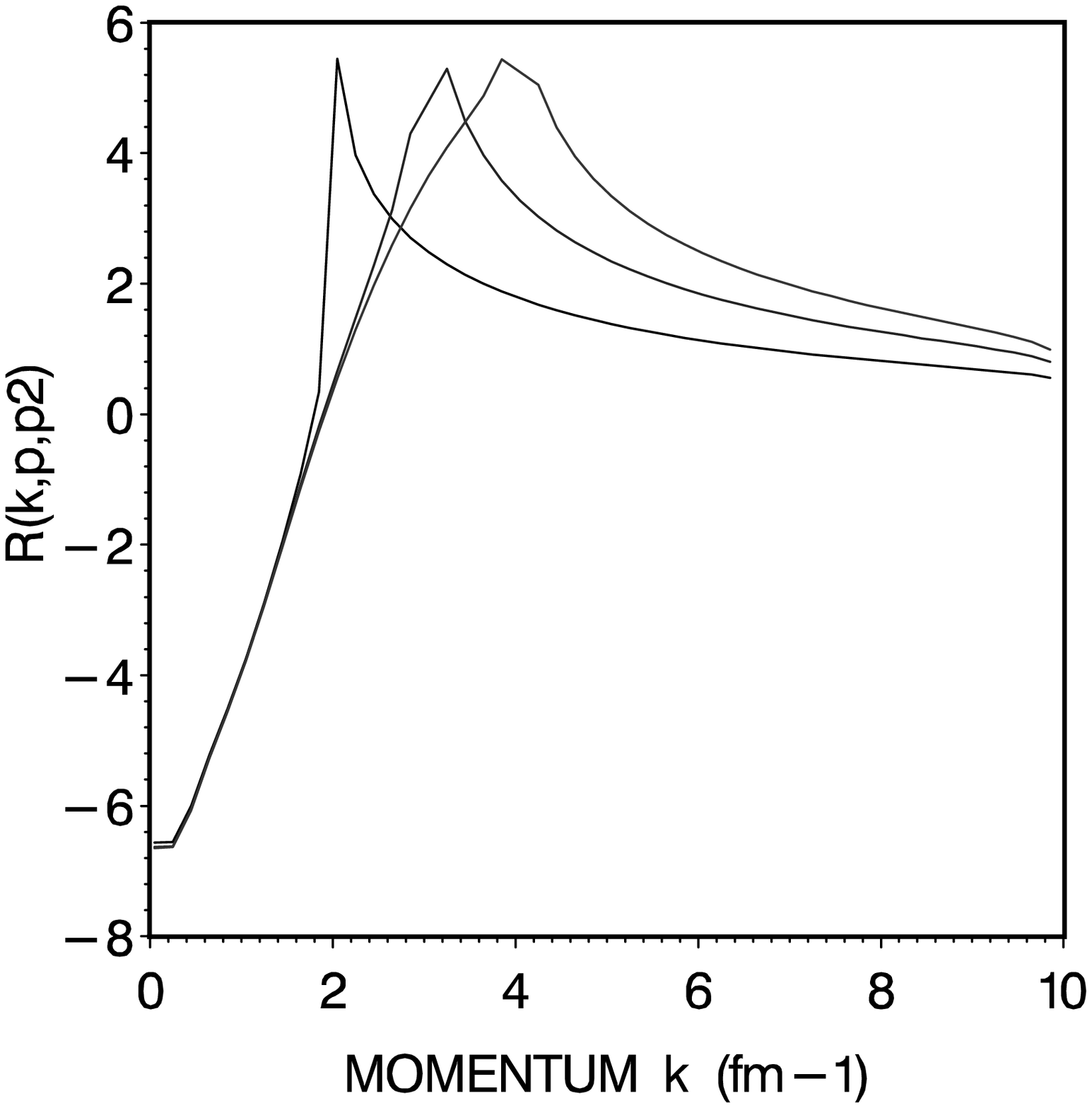,width=7cm,angle=0}
}
\vspace{.0in}
\caption{
Half-shell reactance matrix elements for three different values of
$\Lambda$ and short-ranged parameter $r_c$ as described in the text, with
$\Lambda=2,3$ and $4 fm^{-1}$ and $r_c=0.49,0.68$ and $0.78 fm$ from left to right.
The momentum $p=1 fm^{-1}$.
}
\label{trit8}
\end{figure}

Fig. \ref{trit8} shows the half-shell reactance matrices for the three
values of $\Lambda$. It shows results similar to those of Fig.\ref{trit7}.
It is important to realise that had the  repulsive force  also been cut, one  would
have $R(k,p,p^2)\equiv 0$ for $k>\Lambda$. 

\section{Summary and Discussion}
Separable NN-potentials were used in Triton calculations, with emphasis on
off-shell dependence. The on-shell properties 
were determined from Deuteron data and
scattering phase-shifts by inverse scattering. It seems reasonable to
assume that it would only be necessary to consider low-energy phase-shifts 
as long as energies compatible with those expected in the Triton  are
included. This certainly imposes a necessary condition. But it is not
sufficient. A theory of a many-body system requires also off-shell input. 
The present calculations find that the momentum-range has to be raised
appreciably to adequately include the off-shell scatterings. 
Figs \ref{trit6} to \ref{trit8} illustrates this situation.

Many past results with phase-shift equivalent potentials have ascribed
differences between experimental and microscopically calculated energies as
being alternatively due to off-shell or three-body contributions. The
'equivalence' thorem has however changed this picture. Two-body off-shell
is not observable and off-shell effects and three-body contributions can
not be uniquely separated other than in relation to a specific
potential-model. 

Experiments only provide on-shell data with some off-shell information from
the Deuteron. Off-shell information can be obtained from theoretical input
but specific only to that chosen input. With a 2-body potential so defined, the
calculation of the Triton would  then give information on the three-body
force required together with that specific two-body force.

It was argued above that the $S$ state potentials  are separable for low
momenta as in UPA. This implies that  the
off-shell is also defined for low momenta without any additional 
theory other than that of the importance of a pole in the $T$-matrix. 

It was illustrated above that the  vanishing of the off-shell scatterings 
with decreasing $\Lambda$ could be corrected for by adding a 2-body
short-ranged (contact) term to the attractive long-ranged part.
This term has the effect of increasing correlations and will by eq.
(\ref{domega2}) add a repulsive  component to the long-ranged
attractive \it in-medium \rm interaction, which is of importance in a low
density medium.

Some 'effective' in-medium forces have a similar structure but with a
\it density-dependent \rm 2-body term, e.g. induced  by a 3-body contact force.
One example is the Skyrme-force. Another would be $V_{low - k}$ plus a
three-body force. \cite{dal09} Our effective force represented by 
the $K-$matrix has a different in-medium (density) dependence in that both
the long- and the short-ranged parts  are medium-dependent as shown by eq.
(\ref{G}), with V(k,p) given by eq. (\ref{hhgg}).
Our effective force reduces to the $R-$matrix at zero density irrespective of the
value of $\Lambda$.\footnote{With the limit taken properly to $\delta$
rather than to $tan\delta$\cite{ries56}}.  A $V_{low - k}$ would also do so 
but only for small enough cut-offs $\lambda$.

The model forces (functions of the cut-off) that are a result of this report
are fitted to Triton binding energy and $^2a$ scattering length. As
mentioned above, these forces lead to under-binding of nuclear matter in Brueckner
theory. This may not be significant because  unlike the Triton calculation
it is not exact. Of more interest would e.g. be $^4He$. 

An unanswered question is why the separable (non-local) potential that agreed with Bonn-potential results for $S$-state
Brueckner calculations overbinds the Triton while, as is well-known, realistic (local) potentials in general 
underbind.
It is however also been known for many years and shown by   separate authors that separable potentials fitted to 
low energy phase-shifts (although not from inverse scattering) overbind. The answer may again lie in
differences in off-shell properties but a closer investigation is justified. It is however also believed that
a greater significance should be given to the "exact" Triton rather than to approximate nuclear matter calculations.

\section{Acknowledgements}
It is a pleasure to thank  Prof Bira van Kolck for help in steering me right in
understanding some of the intricacies of EFT and for critical comments on an earlier version of the
manuscript.
I am also grateful to Dr Sid Coon for a careful reading of the manuscript with
helpful comments and suggestions.
Discussions with Prof Nai Kwong were as usual always helpful.

\end{document}